\documentclass[aip,jcp,reprint,twocolumn,superscriptaddress,showpacs,floatfix]{revtex4-2}
\usepackage{mathrsfs}
\usepackage{amssymb}
\usepackage{amsmath}
\usepackage[nodisplayskipstretch]{setspace}
\usepackage{parskip}
\usepackage{graphicx}  % Include the graphicx package
\usepackage{braket}
\usepackage{adjustbox}
\usepackage[dvipsnames]{xcolor}
\usepackage{multirow}
\usepackage{listings}
\usepackage{xcolor}
\usepackage{booktabs}
\usepackage{comment}
\usepackage{graphicx}
\usepackage{lmodern}
\usepackage{subcaption}
\usepackage{float}
\usepackage[normalem]{ulem}
\usepackage[colorlinks=true,citecolor=BrickRed]{hyperref}

\begin{document}
\title{Second-Order Polarization Propagator Approximation to the Quadratic Unitary Coupled Cluster Framework I: First-Order Properties}

\author{Kamal Majee}
\thanks{Contributed Equally.}
\affiliation{Department of Chemistry, Indian Institute of Technology Bombay, Powai, Mumbai 400076, India}

\author{Sudipta Chakraborty}
\thanks{Contributed Equally.}
\affiliation{Department of Chemistry, Indian Institute of Technology Bombay, Powai, Mumbai 400076, India}

\author{Amrita Manna}
\affiliation{Department of Chemistry, Indian Institute of Technology Bombay, Powai, Mumbai 400076, India}
\author{J\'{a}n \v {S}imunek}
\affiliation{ Department of Inorganic Chemistry, Faculty of Natural Sciences, Comenius University, Bratislava
Ilkovičova 6, Mlynská dolina 842 15 Bratislava, Slovakia }
\author{Jozef  Noga}
\affiliation{ Department of Inorganic Chemistry, Faculty of Natural Sciences, Comenius University, Bratislava
Ilkovičova 6, Mlynská dolina 842 15 Bratislava, Slovakia }

\author{Achintya Kumar Dutta}
\thanks{Corresponding author}
\email[e-mail: ]{achintya@chem.iitb.ac.in}
%\email[e-mail: ]{achintya.kumar.dutta@uniba.sk}
\affiliation{Department of Chemistry, Indian Institute of Technology Bombay, Powai, Mumbai 400076, India}
\affiliation{ Department of Inorganic Chemistry, Faculty of Natural Sciences, Comenius University, Bratislava
Ilkovičova 6, Mlynská dolina 842 15 Bratislava, Slovakia }

\begin{abstract}
We have developed a new formulation for calculating first-order properties in the quadratic unitary coupled-cluster singles and doubles (qUCCSD) method, using a second-order polarizable propagator approximation.
The accuracy and reliability of this new framework of the proposed approach have been assessed through the calculation of ground- and excited-state dipole moments and oscillator strengths for a set of small molecules. The Hermiticity of the UCC similarity-transformed Hamiltonian allows molecular properties to be evaluated without introducing left-hand eigenvectors, thereby simplifying the theoretical formulation and computational procedure. Furthermore, the proposed formulation preserves the size-intensivity of transition moments, a property that is not generally guaranteed within the conventional non-Hermitian equation-of-motion coupled-cluster framework. To assess the performance of the proposed method, we have benchmarked both ground- and excited-state properties for 11 small molecules against the extrapolated full configuration interaction (exFCI) reference values and compared with the conventional coupled cluster singles and doubles (CCSD) and third-order algebraic diagrammatic construction (ADC(3)) results. The qUCCSD method yields average errors comparable to those of CCSD while demonstrating significantly improved accuracy over the ADC(3) method.
\end{abstract}

\maketitle

\section{\label{sec:level1}Introduction\protect\\}
Excited-state properties such as dipole moments, transition intensities etc. play a central role in understanding spectroscopic phenomena, photochemical processes, and energy transfer mechanisms, making their accurate theoretical prediction is essential for the characterization of electronically excited states.\cite{cheam1984transition,helgaker2012recent,neese2009prediction,wilson2013methods,norman2018principles,Klessinger1995ExcitedSA,khomskii2014transition,domcke1997theory,Haglund2007} 
Wave-function-based many-body methods can be improved systematically, but excited states are harder to describe than the ground state: they are more sensitive to electron correlation and near-degeneracies, and several electronic configurations often have to be treated on an equal footing. The first-order properties corresponding to a particular state can be evaluated in two different ways: either as expectation values of the corresponding operators or as energy derivatives with respect to an external perturbation associated with the observable. These two formulations are equivalent when the Hellmann-Feynman \cite {hellmann1937einfuhrung,feynman1939forces} theorem is satisfied; however, they may differ, particularly for approximate wave functions. \\
The accuracy of the calculated first-order properties depends upon the underlying electronic structure method.   Coupled cluster theory (CC) \cite{cizek1967advances,mukherjee1989use,bartlett1978many,bartlett2007coupled,crawford2007introduction,shavitt2009many} has long been regarded as the gold standard for ground-state electronic structure calculations due to its size extensivity, systematic improvability, and favorable accuracy–cost balance. Extensions of CC theory to excited states, such as equation-of-motion coupled cluster (EOM-CC)\cite{roweEquationsofMotionMethodExtended1968,stanton1993equation, EMRICH1981379,comeau1993equation} and linear response coupled cluster (LR-CC)\cite{kochCoupledClusterResponse1990,monkhorstCalculationPropertiesCoupledcluster1977, Mukherjee1979AFormalism,sudiptalrcc} formalisms, have been very successful for single-reference systems. However, these approaches rely on a linear parameterization of excitation operators and are based on the non-Hermitian similarity-transformed Hamiltonian. \cite{dutta2017simple,schirmer2010review}  The non-Hermitian nature of the CC ansatz makes the first-order property calculations using analytic derivatives at least twice as expensive as that of the energy calculation. In the expectation-value coupled-cluster (XCC) formalism, the energy functional gives rise to a nonterminating expansion that requires an appropriate truncation scheme. Moreover, owing to the nonvariational nature of conventional truncated coupled-cluster theory, first-order properties evaluated using a perturbatively truncated expectation-value expression generally differ from those obtained through analytic energy derivatives.\cite{bartlett1988expectation}  In addition, evaluating excited-state expectation values requires solving the corresponding left and right eigenvalue problems. This approach is commonly used in equation-of-motion coupled-cluster (EOM-CC) theory, where additional sets of amplitudes are required, especially in derivative-based formulations. One of the formal problems of the EOM-CC expectation value approach is that the transition properties are not size-intensive\textcolor{red}{.}\cite{Kochlr90} \\
In recent years, unitary coupled cluster (UCC)\cite{Kutzelnigg1977, PalUCC84, Hoffmannucc,bartlett1988expectation} theory has emerged as an attractive alternative to conventional CC formulations. By employing an anti-Hermitian cluster operator, UCC leads to a Hermitian similarity-transformed Hamiltonian, which can be advantageous in situations where conventional CC has difficulties. \cite{quccsd2_a,thomas2021complex,hattig2005structure,kohn2007can,kjonstad2017crossing} Also, UCC has been extensively explored in recent years due to its application in quantum simulation and variational quantum algorithms\textcolor{red}{;}\cite{mcardle2020quantum,cao2019quantum,bauer2020quantum} however, its application on classical computers remains rather limited.
A major problem of the UCC ansatz in classical computers is that the similarity-transformed Hamiltonian does not terminate naturally. As a result, practical implementations require an explicit truncation of the expansion, which can compromise the size extensivity of the resulting energy. \cite{pal1982certain} To overcome this issue, Bartlett and co-workers proposed a perturbative truncation of the UCC energy functional denoted as UCC$(n)$, where $n$ specifies the perturbation order. \cite{NogaUCC_a} This scheme retains the size extensivity of the energy and can be systematically extended to include triple and quadruple excitation operators.\cite{BARTLETT1990513} However, this approximation often fails due to the poor convergence behavior of the underlying low-order M\o ller-Plesset (MP) perturbation, particularly for larger and complex molecular systems. \cite{phillips2025comparing}
Recent developments introduced a truncation of the UCC expansion by limiting the rank of the nested commutators using the Bernoulli expansion of the similarity-transformed Hamiltonian.\cite{liu2018unitary} This has demonstrated improved performance in systems where the MP series exhibits poor convergence.\cite{phillips2025comparing} Recently\textcolor{red}{,} Liu \textit{et al.} proposed an alternative to UCC(n) based on a perturbative truncation of the Bernoulli expansion, referred to as UCCn, which is closely related to the ADC hierarchy.\cite{liu2018unitary,hodecker2022theoretical,ChengUCC3_a} Despite these attractive features, reports of property calculations with UCC-based methods remain limited. Bartlett and co-workers have reported an analytical gradient technique for the UCC(4) method. \cite{WATTS1989359} 
Dreuw and co-workers have recently reported the ground and excited-state dipole moments using the UCC3 method. \cite{hodecker2020unitary} Although UCC3 can be used for property calculations, it often gives less accurate results than the standard CCSD method\cite{majee2025relativistic}. DePrince and co-workers found that the commutator-based truncation approach of UCC reaches the full configuration interaction (FCI) limit more quickly and produces more accurate results than methods that rely on perturbative truncation.\cite{phillips2025comparing} Cheng and co-workers introduced a commutator-based, non-perturbative quadratic unitary coupled cluster method with single and double excitations (qUCCSD), demonstrating accuracy comparable to the conventional CC approach for energy calculations\textcolor{red}{.}\cite{qUCCSD1_a} We have extended this formalism to the relativistic regime and computed ground-state correlation energies and properties using both UCC3 and qUCCSD \cite{majee2024reduced,majee2025relativistic}. The qUCCSD method has shown significant improvement over UCC3 for ground-state property calculations owing to a more complete treatment of single excitations\cite{majee2025relativistic}. Perturbative triples corrections to the unitary coupled cluster method have also been developed.\cite{bartlettuccgold,majee2025perturbativetriplescorrectionrelativistic}
The qUCCSD method extends the standard UCC framework by explicitly retaining terms quadratic in the cluster operator in the amplitude equations.  Although qUCCSD has shown promising performance for ground-state energies and properties,\cite{liu2022quadratic,majee2024reduced} excited-state properties have not yet been studied within this framework. \\
In addition to the analytic derivative and expectation value approaches, there exists a third approach for property calculation using 
the polarization propagator approximation.\cite{nielsen1980transition,oddershede1987propagator,bak2000atomic} The  
second-order polarization propagator approximation (SOPPA) has been successfully used in the literature for property calculation\cite{Jorgensen1980SOPPA,Sauer2006QuadraticSOPPA,Kjaer2010NMRBenchmark}. The use of coupled cluster-based propagators, where coupled cluster amplitudes and eigenvectors are used with the second (or higher) order approximation, has been found to be an accurate and inexpensive approach for the calculation of properties\cite{Korona2006TICCSD}. The algebraic diagrammatic construction (ADC)\cite{schirmerRandomphaseApproximationNew1982,trofimov1999consistent,schirmer2018many,dreuwAlgebraicDiagrammaticConstruction2015} can be derived as a propagator approximation to unitary coupled cluster\cite{prasad1980alternative}. Using the same analogy as the coupled cluster propagator, the qUCCSD method can be combined with the second-order intermediate state representation (ISR)\cite{schirmer1991closed,schirmer2004intermediate,schirmer2018many,dreuwAlgebraicDiagrammaticConstruction2015} as a simple way to calculate ground- and excited-state properties.  The aim of this paper is to calculate the ground and excited-state first-order properties using the second-order polarization propagator approximation of the qUCCSD method.
The paper is organized as follows: Section II provides the theoretical description of the method, Section III presents the computational details, and Section IV provides the results and discussion for the selected systems. Finally, conclusions are summarized in Section V.
\section{Theoretical framework}
\subsection{Quadratic unitary coupled-cluster ground-state ansatz}
In unitary coupled-cluster (UCC) theory,\cite{Kutzelnigg1977, PalUCC84, Hoffmannucc} the electronic ground-state wave function is written using a unitary exponential parametrization
\begin{equation}
\label{eq:ucc_wf}
\ket{\Psi_{\mathrm{0}}} = e^{\hat{\sigma}} \ket{\Phi_0},
\end{equation}
where $\ket{\Phi_0}$ is a single-determinant reference wave function and the cluster operator $\hat{\sigma}$ is anti-Hermitian, 
\begin{equation}
\hat{\sigma} = \hat{T} - \hat{T}^\dagger .
\end{equation}
Here $\hat{T}$ and $\hat{T^{\dagger}}$ consist of particle-hole excitation and de-excitation operators, respectively. In the present work, we restrict $\hat{\sigma}$ to single and double excitations,
\begin{equation}
\hat{\sigma} = \hat{\sigma}_1 + \hat{\sigma}_2 ,
\end{equation}
with
\begin{align}
\hat{\sigma}_1 &=
\sum_{ai} t_i^a \{ \hat{a}_a^\dagger \hat{a}_i \}
-\sum_{ai} (t_i^a)^* \{ \hat{a}_i^\dagger \hat{a}_a \}, \\
\hat{\sigma}_2 &=
\frac{1}{4}
\sum_{abij} t_{ij}^{ab}
\{ \hat{a}_a^\dagger \hat{a}_b^\dagger \hat{a}_j \hat{a}_i \}
-
\frac{1}{4}
\sum_{abij} (t_{ij}^{ab})^*
\{ \hat{a}_i^\dagger \hat{a}_j^\dagger \hat{a}_b \hat{a}_a \}.
\end{align}
Occupied and virtual orbitals are denoted by indices $i,j,\ldots$ and $a,b,\ldots$, respectively.
This anti-Hermitian, $(\hat{\sigma}=-\hat{\sigma}^{\dagger})$ parametrization guarantees that the similarity-transformed Hamiltonian
\begin{equation}
\label{eq:Hbar}
\bar{H} = e^{-\hat{\sigma}} \hat{H} e^{\hat{\sigma}}
\end{equation}
is Hermitian.

The ground-state energy and cluster amplitudes are determined from the projected equations,
\begin{align}
\label{eq:ucc_energy}
\braket{\Phi_0 | \bar{H} | \Phi_0} &= E_{\mathrm{gr}}, \\
\label{eq:ucc_amplitude}
\braket{\Phi_\mu | \bar{H} | \Phi_0} &= 0,
\end{align}
where $\ket{\Phi_\mu}$ denotes singly and doubly excited determinants.
\subsection{Commutator expansion and quadratic truncation}
In contrast to conventional coupled-cluster theory, the commutator expansion of the UCC similarity-transformed Hamiltonian Eq.~\ref{eq:Hbar} does not terminate naturally. To construct a practically tractable method, we adopt a truncation strategy based on the commutator order. Using the Bernoulli-number expansion,\cite{KutzlniggBernouli1_a,KutzlniggBernouli2_a,ChengUCC3_a} the transformed Hamiltonian may be written as
\begin{equation}
\label{eq:Hbar_expansion}
\bar{H} = \bar{H}^{(0)} + \bar{H}^{(1)} + \bar{H}^{(2)} + \bar{H}^{(3)} + \cdots ,
\end{equation}
where $\bar{H}^{(n)}$ gathers all terms that contain exactly $n$ nested commutators between the Hamiltonian and $\hat{\sigma}$.

Truncating Eq.~\eqref{eq:ucc_energy} up to the third commutator and Eq.~\eqref{eq:ucc_amplitude} up to the second commutator leads to quadratic unitary coupled-cluster singles and doubles (qUCCSD) approximation.\cite{qUCCSD1_a,quccsd2_a} 
\begin{align}
\label{eq:quccsd_energy}
\braket{\Phi_0 | \bar{H}^{(0)} + \bar{H}^{(1)} + \bar{H}^{(2)} + \bar{H}^{(3)} | \Phi_0} &= E_{\mathrm{gr}}, \\
\label{eq:quccsd_amplitude}
\braket{\Phi_\mu | \bar{H}^{(0)} + \bar{H}^{(1)} + \bar{H}^{(2)} | \Phi_0} &= 0,
\end{align}
With this truncation, orbital relaxation and correlation effects are described consistently.\cite{majee2024reduced}
\subsection{Excited states from the UCC-based polarization propagator}
Excited states are described using the self-consistent polarization propagator formulation, in which excitation energies are obtained as eigenvalues of the Hermitian-transformed Hamiltonian in an approximate many-electron excitation manifold.\cite{liu2018unitary,prasadAspectsSelfconsistentPropagator1985} \\
The excited-state wave functions are expressed as
\begin{equation}
\label{eq:excited_wf}
\ket{\Psi_L} =
\sum_I C_{IL} \,
e^{\hat{\sigma}} \hat{\tau}_I^\dagger \ket{\Phi_0},
\end{equation}
where $\{\hat{\tau}_I^\dagger \}$ denotes a set of bare excitation operators spanning single and double excitations, which can be defined as
\begin{equation}
\label{eq:excited_operator}
\{ \hat{\tau}_I^\dagger \}=\left\{\hat{a}_a^{\dagger}\hat{a}_i\right\}\cup\left\{
\hat{a}_a^{\dagger}\hat{a}_b^{\dagger}\hat{a}_j\hat{a}_i
\right\}
\end{equation}

The corresponding secular equation takes the form
\begin{equation}
\label{eq:secular}
\sum_I \bar{H}_{JI} C_{IL} = E_L C_{JL},
\end{equation}
with matrix elements
\begin{equation}
\label{eq:Hbar_matrix}
\bar{H}_{JI} =
\braket{\Phi_0 | \hat{\tau}_{_J} \bar{H} \hat{\tau}_{_I}^\dagger | \Phi_0}.
\end{equation}
Because $\bar{H}$ is Hermitian, the eigenvalues $E_{L}$ (and hence the excitation energies) obtained from Eq.~(\ref{eq:secular}) are real.

\subsection{Balanced truncation of the excited-state secular matrix}
The excited-state secular matrix equation can be partitioned into blocks corresponding to the single- and double-excitation spaces, giving the eigenvalue equation, expressed by,
\begin{equation}
\renewcommand{\arraystretch}{1.5}
\begin{bmatrix}
\bar{H}_{SS} & \bar{H}_{SD} \\ \
\bar{H}_{DS} & \bar{H}_{DD}
\end{bmatrix}
\begin{bmatrix}
C_S \\ \
C_D
\end{bmatrix}
=
E
\begin{bmatrix}
C_S \\ \
C_D
\end{bmatrix}
\label{eq:block_matrix}
\end{equation}
Where $\bar{H}_{SS}$, $\bar{H}_{SD}$, $\bar{H}_{DS}$, and $\bar{H}_{DD}$ represent the singles-singles, singles-doubles, doubles-singles, and doubles-doubles blocks, respectively. Within the Löwdin partitioning framework, the effect of the double excitations is incorporated into the single-excitation space.
\begin{equation}
[\bar{H}_{SS}+\bar{H}_{SD}\left(E - \bar{H}_{DD} \right)^{-1}\bar{H}_{DS}]C_S=E\, C_S
\end{equation}
In this partitioning approach, the accuracy is predominantly determined by the $\bar{H}_{SS}$ block, whereas the other remaining blocks, involving double excitation, act as a correction to the singles-singles block. In the qUCCSD scheme,\cite{quccsd2_a} a balanced commutator truncation is employed to ensure uniform accuracy across the different blocks. Specifically, the $\bar{H}_{SS}$ is truncated up to double commutators, the $\bar{H}_{SD}$ and $\bar{H}_{DS}$ blocks are truncated up to single commutators, and the $\bar{H}_{DD}$ block is retained at the level of the bare Hamiltonian. The qUCCSD method extends the UCC3 approach by including higher-order contributions in the cluster-amplitude equations as well as the excited-state secular equations. Thus, UCC3 can be regarded as a subset of the qUCCSD framework, while qUCCSD offers a systematic improvement over UCC3.

\subsection{Polarization propagator approach for UCC representation}

The UCC scheme can be viewed as a reformulation of the diagrammatic perturbation expansion of the polarization propagator.\cite{prasadAspectsSelfconsistentPropagator1985,liu2018unitary,schirmerRandomphaseApproximationNew1982}
We consider an $N$-electron system described by the Hamiltonian $\hat{H}$, with a non-degenerate ground state $|\Psi_0\rangle$ and corresponding ground-state energy $E_0$.

The polarization propagator is defined as
\begin{equation}
\Pi(\omega) = \Pi^{+}(\omega) + \Pi^{-}(\omega),
\label{eq:Pi_def}
\end{equation}
where the matrix elements of the resolvent operator define the forward component,
\begin{equation}
\Pi^{+}_{rs,r's'}(\omega)
=
\left\langle \Psi_0 \left|
\hat{c}_s^\dagger \hat{c}_r
\left( \omega - \hat{H} + E_0 + i\eta \right)^{-1}
\hat{c}_{r'}^\dagger \hat{c}_{s'}
\right| \Psi_0 \right\rangle',
\label{eq:Pi_plus}
\end{equation}
and
\begin{equation}
\Pi^{-}_{rs,r's'}(\omega)
=
\Pi^{+}_{s'r',sr}(-\omega).
\label{eq:Pi_minus}
\end{equation}

Here, $\hat{c}_p^\dagger$ and $\hat{c}_p$ denote the usual creation and annihilation operators. 
The prime $'$ in Eq.~(\ref{eq:Pi_plus}) indicates the subtraction of the disconnected contribution
\begin{equation}
\langle \Psi_0 | \hat{c}_s^\dagger \hat{c}_r | \Psi_0 \rangle
\langle \Psi_0 | \hat{c}_{r'}^\dagger \hat{c}_{s'} | \Psi_0 \rangle
(\omega + i\eta)^{-1}.
\end{equation}
The positive infinitesimal $\eta$ ensures the proper analytic structure of the propagator and will be omitted where unnecessary.

Since $\Pi^{+}(\omega)$ and $\Pi^{-}(\omega)$ contain equivalent physical information, it is sufficient to consider only the forward component $\Pi^{+}(\omega)$.

In the ADC approach, the polarization propagator is written as
\begin{equation}
\Pi^{+}(\omega)
=
\mathbf{f}^{\dagger}
(\omega - \mathbf{M})^{-1}
\mathbf{f},
\label{eq:ADC_form}
\end{equation}
where $\mathbf{M}$ is the secular matrix representation of the shifted Hamiltonian in the intermediate-state basis $|\tilde{\Psi}_{J} \rangle$, defined as
\begin{equation}
    M_{IJ}=\langle \tilde{\Psi}_{I} | \hat{H}-E_{0} | \tilde{\Psi}_{J} \rangle,
\end{equation}
and $\mathbf{f}$ is the corresponding vector of transition amplitudes represented as
\begin{equation}
    f_{J,rs}=\langle \tilde{\Psi}_{J} |\hat{c}_{r}^{\dagger}\hat{c}_{s}|\Psi_{0} \rangle.
    \label{eq:effective}
\end{equation}
\begin{comment}
The excitation energies are obtained from the eigenvalue problem
\begin{equation}
\mathbf{M}\mathbf{X} = \mathbf{X}\boldsymbol{\Omega}, 
\qquad 
\mathbf{X}^\dagger \mathbf{X} = \mathbf{1},
\label{eq:pp}
\end{equation}
where $\boldsymbol{\Omega}$ the diagonal matrix of excitation energies, and $\mathbf{X}$ contains the corresponding eigenvectors $\mathbf{X}_n$ as its columns. \\
In the intermediate state representation (ISR), the $\mathbf{M}$ and $\mathbf{f}$ components are affordably constructed directly. In this scheme, the Hamiltonian is shifted by the ground-state energy $E_{0}$ expressed within the orthonormal basis, so-called intermediate states (ISs)
\begin{equation}
    M_{IJ}=\langle \tilde{\Psi}_{I} | \hat{H}-E_{0} | \tilde{\Psi}_{J} \rangle ,
\label{eq:intermediate_state}
\end{equation} \\
\end{comment}
%%%%%%%%%%%%%%%%%%%%%
In the UCC approach, excited-state wave functions are generated from the ground-state UCC wave function by using a so-called self-consistent operator $(e^{\hat{\sigma}}\hat{C}_{J}e^{-\hat{\sigma}})$, 
\begin{equation}
\label{eq:excited_state}
\vert\tilde{\Psi}_{J} \rangle = e^{\hat{\sigma}}\hat{C}_{J}e^{-\hat{\sigma}}\vert\Psi_{0} \rangle = e^{\hat{\sigma}}\hat{C}_{J}\vert \Phi_{0} \rangle
\end{equation}
where $\hat{C}_{J}$ denotes ``physical" excitation operators of particle-hole (p-h), (2p-2h), $\cdots$ type, represented as 
\begin{equation}
\label{eq:17}
\{\hat{C}_J\} = \{\hat{c}_{a}^{\dagger}\hat{c}_{i};\hat{c}_{a}^{\dagger}\hat{c}_{b}^{\dagger}\hat{c}_{j}\hat{c}_{i}, a<b,i<j,\cdot\cdot\cdot \}
\end{equation}
The excitation energies are obtained by representing the shifted Hamiltonian $(\hat{H}-E_{0})$ in the orthonormal basis,
\begin{align}
\label{eq:18}
\tilde{M}_{IJ}
&=
\langle \tilde{\Psi}_I \vert \hat{H}-E_{0} \vert \tilde{\Psi}_J \rangle
\nonumber \\
&=
\langle \Phi_{I} \vert \bar{H} \vert \Phi_{J} \rangle -E_{0}\delta_{IJ},
\end{align}
so diagonalizing the Hermitian eigenvalue equation
\begin{equation}
\tilde{\mathbf{M}}\mathbf{X} = \mathbf{X}\boldsymbol{\Omega}, 
\qquad 
\mathbf{X}^\dagger \mathbf{X} = 1
\label{eq:pp}
\end{equation}
yields vertical excitation energies as eigenvalues $\Omega_{n}=E_{n}-E_{0}$, collected in the diagonal matrix $\boldsymbol{\Omega}$, and the corresponding excitation vectors as eigenvectors collected in \textbf{X}.

\subsection{Second-order polarization propagator approach to qUCCSD for Ground and excited-state molecular properties}
\label{sec:quccsd2}
In the present work, we extend this second-order polarization-propagator framework to the qUCCSD method, abbreviated as qUCCSD(2), for the calculation of the ground- and excited-state molecular properties. The key idea of the qUCCSD(2) approximation is to use the qUCCSD ground- and excited-state wave functions combined with a second-order ISR expression derived within the polarization propagator framework. \\
In the ISR space $\tilde{\mathbf{M}}$ is partitioned into 1p-1h and 2p-2h blocks. The corresponding secular matrix is therefore written in block form as 
\begin{equation}
\tilde{\mathbf{M}}
=
\begin{pmatrix}
{\mathbf{\tilde{M}}_{11}} & \mathbf{\tilde{M}_{12}}\\
\mathbf{\tilde{M}_{21}} & \mathbf{\tilde{M}_{22}}
\end{pmatrix},
\end{equation}
where the indices 1 and 2 refer to the 1p-1h and 2p-2h intermediate-state subspace, respectively.
The qUCCSD(2) approximation is introduced here, applying a block-dependent truncation to the BCH expansion of the similarity-transformed Hamiltonian. Specifically, the $\mathbf{\tilde{M}_{11}}$ block is evaluated by retaining contributions up to the second nested commutator, whereas the off-diagonal coupling blocks $\mathbf{\tilde{M}_{12}}$ and $\mathbf{\tilde{M}_{21}}$ are retained up to the first commutator, and the $\mathbf{\tilde{M}_{22}}$ block is restricted to the bare Hamiltonian contribution. According to the $\mathbf{\tilde{M}}$ defined as,
\begin{equation}
\mathbf{\tilde{M}}
=
\begin{pmatrix}
\mathbf{\tilde{M}_{11}^{[2]}} & \mathbf{\tilde{M}_{12}^{[1]}}\\
\mathbf{\tilde{M}_{21}^{[1]}} & \mathbf{\tilde{M}_{22}^{[0]}}
\end{pmatrix},
\label{eq:quccsd_secular_matrix}
\end{equation}
Where the subscript $[n]$ denotes the highest order of nested commutator retained in the BCH expansion. Therefore, the matrix structures of qUCCSD and ADC(2) are analogous.\\
To evaluate the molecular properties, we introduce a general one-particle operator $\hat{D}$.
The excited-state expectation value can be written as,
\begin{equation}
    D_{n}=\langle \Psi_{n}|\hat{D}|\Psi_{n}\rangle
\label{eq:expectation_value}
\end{equation}
The excited-state wave function $\ket{\Psi_{n}}$ is expanded in terms of the intermediate states (IS) as
\begin{equation}
    |\Psi_{n}\rangle = \sum_{J} X_{Jn} | \tilde{\Psi}_{J}\rangle,
\label{eq:ex}
\end{equation}
where $X_{Jn}=\langle{\tilde{\Psi}{_J}}|{\Psi_{n}}\rangle$ are the expansion coefficients and $(\tilde{\Psi}_{J})$ are the IS wave functions. \\
The expectation value of $D_{n}$ $(n > 0)$  within the IS basis can be expressed as,
\begin{equation}
D_{n}
=
\boldsymbol{X_{n}^{\dagger}\, \tilde{D}\, X_{n}}.\\
\label{eq:isr_xn_d_xn}
\end{equation}
where $\tilde{D}$ is the intermediate state representation (ISR) of the operator $\hat{D}$ with the elements
\begin{equation}
    \tilde{D}_{IJ}=\langle \tilde{\Psi}_{I} \vert \hat{D} \vert \tilde{\Psi}_{J} \rangle
\end{equation}
The ground-to-excited-state transition moment is
\begin{equation}
    T_{n}=\langle \Psi_{n} \vert \hat{D} \vert \Psi_{0} \rangle
\label{eq:transition_moment}
\end{equation}
where the one-particle operator $\hat{D}$ is represented as
\begin{equation}
    \hat{D}=\sum_{r,s}d_{rs}\hat{c}_{r}^{\dagger}\hat{c}_{s}.
\label{eq:one_particle_operator}
\end{equation}
where, 
\begin{equation}
    d_{rs}=\langle \varphi_{r}|\hat{d}| \varphi_{s} \rangle,
\label{eq:one_particle_integrals}
\end{equation}
represents the one-particle matrix elements associated with $\hat{D}$.
The transition moment in the ISR basis can be formulated as
\begin{equation}
    T_{n}=\sum_{J}X_{Jn}^{*}F_{J},\quad \quad 
\end{equation}
where $F_{J}=\langle \tilde{\Psi}_{J}|\hat{D}|\Psi_{0} \rangle$ defines the ISR transition moment for the operator $\hat{D}$ and can be further expanded as
\begin{equation}
    F_{J}=\sum_{r,s}f_{_{J,rs}}d_{_{rs}},
\end{equation}
where,
\begin{equation}
    f_{_{J,rs}}=\langle \tilde{\Psi}_{J}|\hat{c}_{r}^{\dagger}\hat{c}_{s}|\Psi_{0} \rangle
\label{eq:transition_amplitude}
\end{equation}
are the ISR transition amplitudes.
This ISR transition amplitude can be expanded perturbatively
\begin{equation}
    \textbf{f}=\textbf{f}^{(0)}+\textbf{f}^{(1)}+\textbf{f}^{(2)}+\cdots
\end{equation}
The spectral-amplitude vector is partitioned according to the 1p-1h and 2p-2h intermediate state space. To obtain ground-to-excited-state transition moments through second order, the spectral amplitudes are expanded perturbatively in the same spirit as the ADC(2) property formulation. The 1p-1h component of the spectral-amplitude vector contains contributions through second order, whereas the 2p-2h component is retained through first order. Accordingly, the second-order spectral amplitude vector is written as
\begin{equation}
\mathbf{f}
=
\begin{pmatrix}
\mathbf{f}_{1}^{(0-2)}\\
\mathbf{f}_{2}^{(1)}
\end{pmatrix},
\label{eq:spectral_amplitude}
\end{equation}
For the excited-state property calculation involving two different states ($n \neq m$), one may compute the corresponding transition moment by
\begin{equation}
    T_{nm}=\langle\Psi_{n}\vert \hat{D}\vert \Psi_{m}\rangle = \mathbf{X_{n}^{\dagger}\tilde{D}X_{m}}
\label{eq:transition_moment_nm}
\end{equation}
It is convenient to separate the ground-state contribution from the ISR property matrix. The matrix elements can therefore be written as 
\begin{equation}
    \tilde{D}_{IJ}=D_{0}\delta_{IJ}+D_{IJ}
\end{equation}
where $D_{0}=\langle \Psi_{0} \vert \hat{D} \vert \Psi_{0} \rangle$ is the ground-state expectation value, which contributes only to the diagonal elements.
Therefore, the subtracted operator can be defined as,
\begin{equation}
\boldsymbol{\mathcal{\hat{D}}}=\hat{D}-D_{0} 
\end{equation}
This is analogous to the shifted Hamiltonian, $(\hat{H}-E_{0})$ used in the secular problem. \\
Within the UCC formulation using the definition in Eq.~\ref{eq:excited_state}, one can express the ISR property matrix in terms of the similarity-transformed property operator. Using the definition of the intermediate states, the matrix elements are given by 
\begin{align}
\tilde{D}_{IJ}
&= \langle \Phi_{0} | \hat{C}_{I}^{\dagger} e^{-\sigma} \hat{D} e^{\sigma}
   \hat{C}_{J} | \Phi_{0} \rangle \notag \\
&= \langle \Phi_{I} | \bar{D} | \Phi_{J} \rangle
\label{eq:ISR_property}
\end{align}
where
\begin{equation}
\bar{D} = e^{-\sigma} \hat{D} e^{\sigma}
\end{equation} 
denotes the similarity-transformed operator, and one can expand it by BCH expansion. Thus,
Eq.~\ref{eq:ISR_property} can therefore be obtained from the matrix elements of $\bar{D}$ between the corresponding intermediate configurations, and it can be partitioned according to the same excitation classes as the secular matrix,
\begin{equation}
\mathbf{\tilde{D}}
=
\begin{pmatrix}
\mathbf{\tilde{D}}^{[2]}_{11} &
\mathbf{\tilde{D}}^{[1]}_{12}\\
\mathbf{\tilde{D}}^{[1]}_{21} &
\mathbf{\tilde{D}}^{[0]}_{22}
\end{pmatrix}.
\end{equation}

However, in the present implementation, the truncation of this property matrix is treated using the second-order propagator approximation. The property matrix is constructed by explicitly expanding its individual matrix elements according to their perturbative order, as in the ADC(2) truncation\cite{schirmer2004intermediate}. At the second-order level, the required contributions to the individual blocks are
\begin{equation}
\begin{aligned}
\mathbf{\tilde{D}}_{11}^{(0-2)} &= \mathbf{\tilde{D}}_{11}^{(0)} + \mathbf{\tilde{D}}_{11}^{(1)} + \mathbf{\tilde{D}}_{11}^{(2)}, \\
\mathbf{\tilde{D}}_{12}^{(0-1)} &= \big[\mathbf{\tilde{D}}_{21}^{(0-1)}\big]^{\dagger} = \mathbf{\tilde{D}}_{12}^{(0)} + \mathbf{\tilde{D}}_{12}^{(1)}, \\
\mathbf{\tilde{D}}_{22}^{(0)} &= \mathbf{\tilde{D}}_{22}^{(0)}.
\end{aligned}
\end{equation}
Consequently, the second-order property matrix can be represented schematically as
\begin{equation}
\mathbf{\tilde{D}}
=
\begin{pmatrix}
\mathbf{\tilde{D}}_{11}^{(0-2)} &
\mathbf{\tilde{D}}_{12}^{(0-1)}
\\
\mathbf{\tilde{D}}_{21}^{(0-1)} &
\mathbf{\tilde{D}}_{22}^{(0)}
\end{pmatrix}.
\end{equation}
Here, the superscript in parentheses denotes the perturbative order of the corresponding contribution.
\begin{comment}

\subsection{Relation to existing excited-state methods}
In the absence of beyond-third-order terms up to the quadratic commutator, the present formulation reduces to previously developed perturbative UCC-based polarization propagator approaches.\cite{liu2018unitary,hodecker2020unitary} The explicit inclusion of those beyond-third-order terms upto the quadratic commutator contributions leads to improved robustness and accuracy, particularly for systems exhibiting strong orbital-relaxation effects or slow convergence of M{\o}ller--Plesset perturbation series. The Hermitian nature of the qUCCSD Hamiltonian further enables a physically consistent description of near-degenerate and same-symmetry excited states.
\end{comment}
\section{Computational Details and Implementation}
We have implemented the qUCCSD(2) method for calculating ground- and excited-state properties in our in-house software package, BAGH.~\cite{duttaBAGHQuantumChemistry2025_a} The interface of BAGH depends on some external software packages, including PySCF,\cite{sun2020recent,sun2018pyscf,sun2015libcint} socutils,\cite{socutils} DIRAC,\cite{saue_2025_14833106} and GAMESS-US.\cite{barca2020recent} For this work we have used PySCF interface.
In this paper, we have tested 11 molecules, including BH, HCl, H$_{2}$O, H$_{2}$S, BF, CO, H$_{2}$CO, H$_{2}$CS, HNO, FCH, and H$_{2}$CSi. All molecular geometries and extrapolated FCI (exFCI) data were taken from the work of Damour \textit{et al.}~\cite{damour2022ground}, whereas the CC results were taken from the work of Chrayteh \textit{et al.}~\cite{chrayteh2020mountaineering}. The statistical parameters that has been used to analyze errors are mean absolute error (MAE), mean error (ME), standard deviation (STD), root mean square error (RMSE) and maximum error ($\Delta_{max})$.
\section{Results and Discussion}
\begin{table*}[ht!]
\centering
\caption{Ground- and excited-state dipole moments (D) and length-gauge oscillator strengths at the qUCCSD(2), CCSD, and exFCI levels using the aug-cc-pVTZ basis set. V and R denote valence and Rydberg states, respectively.}
\label{tab:all_dipole_tz}
\begin{ruledtabular}
\begin{tabular}{llc cc cc cc }
 &  &  & \multicolumn{2}{c}{qUCCSD(2)} &\multicolumn{2}{c}{CCSD\textsuperscript{a}} &\multicolumn{2}{c}{exFCI\textsuperscript{a}} \\
\cline{4-5} \cline{6-7}\cline{8-9}
Molecule & Excitation & Nature 
& $|\mu|$ & $f^{L}$ 
& $|\mu|$ & $f^{L}$
& $|\mu|$ & $f^{L}$ \\
\hline

BH  & $^1\Sigma^+$                                  &           &  1.411 &         &1.433   &       & 1.408 &               \\
    & $^1\Pi$                                       & V         &  0.521 & 0.049   &0.550   & 0.050 & 0.554 & 0.048     \\ \\
                
HCl & $^1\Sigma^+$                                  &           &  1.094 &         & 1.097  &       & 1.084 &               \\
    & $^1\Pi$                                       & V         &  2.459 & 0.050   & 2.526  & 0.056 & 2.501 & 0.055     \\ \\
                
H$_2$O & $^1A_1$                                    &           &  1.865 &         & 1.864  &       & 1.840 &           \\
       & $^1B_1(n \rightarrow 3s)$                  & R         &  1.677 & 0.052   & 1.549  & 0.053 & 1.557 & 0.054    \\
       & $^1A_2(n \rightarrow 3p)$                  & R         &  1.249 &         & 1.056  &       & 1.105 &         \\
       & $^1A_1(n \rightarrow 3s)$                  & R         &  1.288 & 0.100   & 1.163  & 0.098 & 1.214 & 0.100   \\ \\
                
H$_2$S & $^1A_1$                                    &           &  0.986 &         & 0.990  &       & 0.977 &         \\
       & $^1A_2(n \rightarrow 4p)$                  & R         &  0.550 &         & 0.503  &       & 0.499 &         \\
       & $^1B_1(n \rightarrow 4s)$                  & R         &  1.868 & 0.061   & 1.893  & 0.064 & 1.866 & 0.063   \\ \\
                
BF  & $^1\Sigma^+$                                       &           &  0.776 &         & 0.794  &       & 0.824 &         \\
    & $^1\Pi(\sigma \rightarrow \pi^*)$             & V         &  0.195 & 0.473   & 0.222  & 0.475 & 0.294 & 0.468   \\ \\
                
CO  & $^1\Sigma^+$                                  &           &  0.055 &         & 0.051  &       & 0.116 &         \\
    & $^1\Pi(n \rightarrow \pi^*)$                  & V         &  0.179 & 0.167   & 0.227  & 0.161 & 0.129 & 0.166   \\ \\
                
H$_2$CO & $^1A_1$                                   &           & 2.445  &         & 2.457  &       & 2.384 &         \\
        & $^1A_2(n \rightarrow \pi^*)$              & V         & 1.258  &         & 1.416  &       & 1.325 &         \\ \\
                
H$_2$CS & $^1A_1$                                   &           & 1.697  &         & 1.737  &       & 1.695 &         \\
        & $^1A_2(n \rightarrow \pi^*)$              & V         & 0.753  &         & 0.848  &       & 0.839 &         \\ \\
                
HNO & $^1A'$                                        &           & 1.714  &         & 1.722  &       & 1.676 &         \\
    & $^1A''(n \rightarrow \pi^*)$                  & V         & 1.665  &         & 1.727  &       & 1.675 &         \\ \\
            
FCH & $^1A'$                                        &           & 1.463  &         & 1.465  &       & 1.439 &         \\
    & $^1A''$                                       & V         & 0.925  & 0.007   & 0.991  & 0.006 & 0.958 & 0.006  \\ \\
    
H$_2$CSi                                & $^1A_1$   &           & 0.240  &        &  0.235  &       & 0.137 &         \\
                                        & $^1A_2$   & R         & 2.112  &        &  1.851  &       & 1.933 &         \\
                                        & $^1B_2$   & R         & 0.110  & 0.036  &  0.161  & 0.036 & 0.041 & 0.034   \\

\end{tabular}
\end{ruledtabular}
\noindent\textsuperscript{a} Values are taken from Ref. \cite{fullCIgsandex}
\end{table*}

Table~\ref{tab:all_dipole_tz} collects ground- and excited-state dipole moments and oscillator strengths computed at the qUCCSD(2), CCSD, and exFCI levels with the aug-cc-pVTZ basis set for a diverse set of small molecules. Overall, qUCCSD(2) agrees well with exFCI, although the quality of agreement varies with the molecule and with the character of the excited state.

\begin{table}[ht]
\centering
\caption{Ground-state dipole moment errors (Debye) with respect to reference values (exFCI) using aug-cc-pVTZ basis set. The sample size (N) is 11.}
\label{tab:gs_dip_errors_dz}
\begin{ruledtabular}
\begin{tabular}{lccccc}
Method     & ME       & MAE   & RMSE & STD   & $\Delta_{max}$ \\
\hline
ADC(3)         & 0.095 & 0.111 & 0.152 & 0.119 & 0.281 \\
CCSD           & 0.024 & 0.041 & 0.049 & 0.042 & 0.098 \\
%qUCCSD(2)  & 0.022 & 0.047 & 0.057 & 0.053 & 0.101 \\
qUCCSD(2) & 0.015& 0.035& 0.046& 0.043&0.103\\
\end{tabular}
\end{ruledtabular}
\end{table}

\subsection{Ground-State Dipole Moments}
Table~\ref{tab:gs_dip_errors_dz} summarizes the errors in the ground-state dipole moments of the 11 molecules with respect to exFCI, and the individual values for all methods, including CCSDT, are collected in Tables S1--S12 of the Supporting Information. qUCCSD(2) reproduces the exFCI dipole moments closely, with an MAE of 0.035~D and an RMSE of 0.046~D. The small positive ME (0.015~D) indicates only a slight tendency to overestimate the dipole moment. For seven of the eleven molecules (BH, HCl, H$_2$O, H$_2$S, H$_2$CS, HNO, and FCH) the deviation from exFCI is below 0.04~D.

Among the four remaining systems, BF and H$_2$CSi show deviations of $-0.048$~D and $+0.103$~D, respectively.CO and H$_2$CO both deviate by $0.061$~D, with an underestimation for CO and an overestimation for H$_2$CO
CCSD underestimates the dipole moments of BF and CO by 0.030 and 0.065~D, respectively, and overestimates those of H$_2$CO and H$_2$CSi by 0.073 and 0.098~D. With CCSDT, the absolute error reduces to 0.016~D or less for all four molecules as shown in Tables S6--S8 and S12.
These deviations therefore reflect the missing connected triple excitations common to both singles-and-doubles methods, rather than a shortcoming specific to the unitary ansatz or to the second-order property expression. For CO and H$_2$CSi, the reference dipole moments are small (0.116 and 0.137~D), so modest absolute errors translate into large relative ones: both qUCCSD(2) (0.055~D) and CCSD (0.051~D) recover less than half of the exFCI dipole moment of CO. The T$_1$ diagnostics (Table S15) remain below 0.03 for all molecules at both the qUCCSD and CCSD levels, which indicates that none of these deviations originates from pronounced multireference character.

qUCCSD(2) gives slightly smaller average errors than CCSD, for which the MAE, RMSE, and ME are 0.041, 0.049, and 0.024~D, respectively.
The standard deviations are nearly identical, at 0.043~D for qUCCSD(2)
and 0.042~D for CCSD, as are the maximum deviations of 0.103 and
0.098~D, respectively. The MAE differs by only 0.006~D, which is small
relative to the spread of the individual errors. Overall, the two
methods provide comparable accuracy for ground-state dipole moments
in this test set. For BH and H$_2$CS, qUCCSD(2) gives errors of
$+0.003$ and $+0.002$~D, respectively, considerably smaller than
the corresponding CCSD errors of $+0.025$ and $+0.042$~D.
CCSD, however, gives a more accurate dipole moment for BF.

ADC(3) gives much larger errors, with an MAE of 0.111~D, an RMSE of 0.152~D, and a maximum deviation of 0.281~D, roughly three times the corresponding qUCCSD(2) values. The positive ME of 0.095~D shows a systematic overestimation. The errors, however, vary strongly between molecules. ADC(3) is within 0.013~D of exFCI for H$_2$O, H$_2$CS, and HNO, but overestimates the dipole moments of BH, BF, and CO by 0.26--0.28~D and that of FCH by 0.12~D. This behavior reflects the perturbative ground state underlying ADC. The MP3 dipole moments (Table S13) overestimate the exFCI values for every molecule, with an MAE of 0.22~D, which shows that a low-order perturbative description of the ground state is not sufficient for this property. In qUCCSD(2), the perturbative amplitudes are replaced by iteratively converged qUCCSD amplitudes, and this change alone lowers the MAE from 0.111 to 0.035~D and the maximum error from 0.281 to 0.103~D.

\subsection{Size Intensivity of Transition Properties}
\begin{figure}[ht!]
    \centering
    \includegraphics[width=0.5\textwidth]{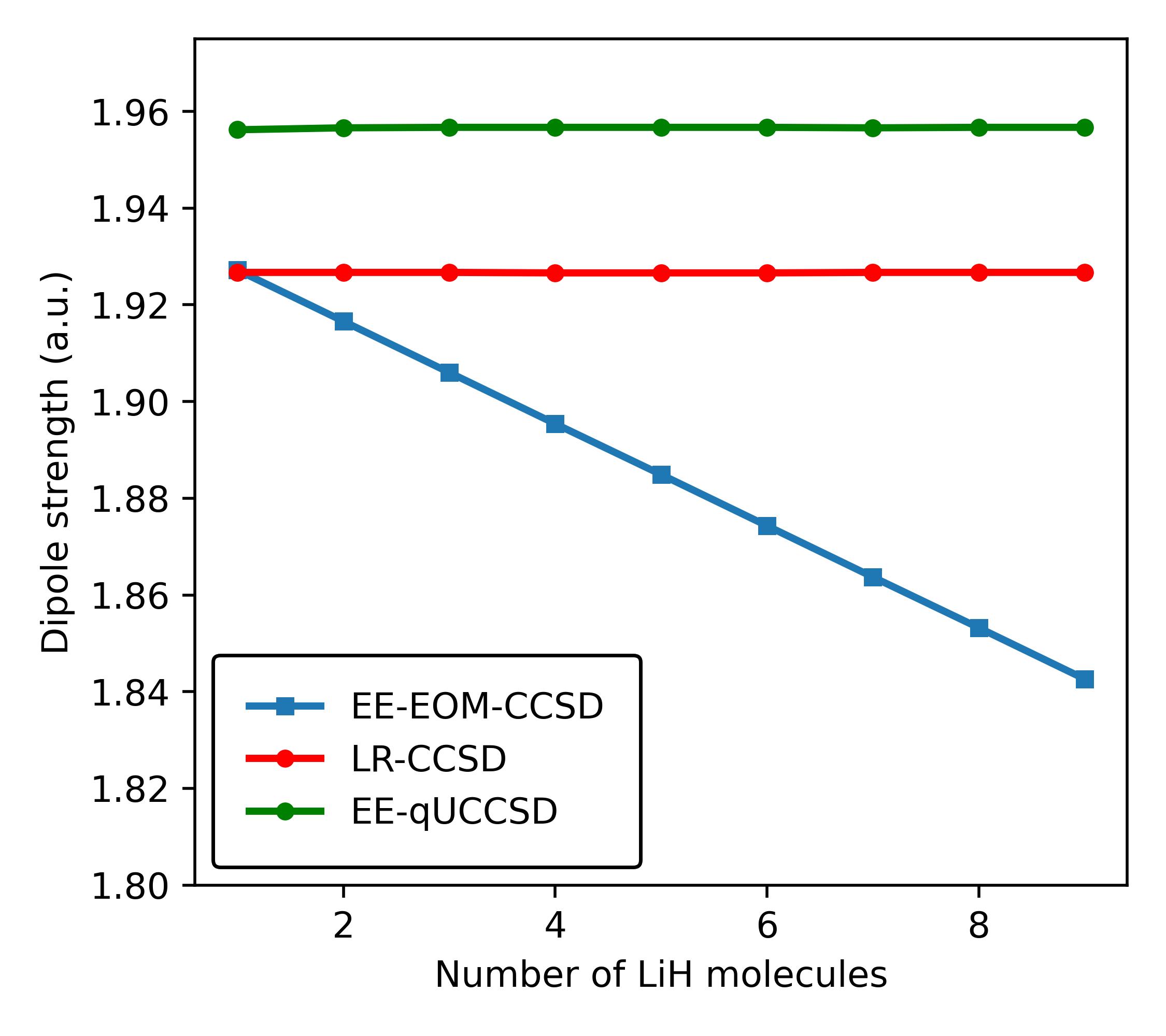} % Replace with your actual path
    % Main caption for the figure
    \caption{Dipole strength (a.u.) as a function of the number of non-interacting LiH molecules.}
    \label{fig:size_extensivity}
\end{figure}
A known shortcoming of EOM-CCSD is that its transition properties are not size-intensive. To assess the size-intensivity of the qUCCSD(2) transition moment, we studied a model system of non-interacting LiH molecules. This system was previously investigated by Koch \textit{et al.},\cite{Kochsize-extensive} where the failure of the EOM-CC method to yield size-intensive properties was first highlighted. 
We repeated the calculations using the qUCCSD(2) method for the same system in aug-cc-pVDZ basis set with a Li–H bond distance of 4 a.u., and extended it along the y-axis with an intermolecular separation of 5000 a.u. Figure~\ref{fig:size_extensivity} shows the dipole strength as a function of the number of LiH molecules, and the corresponding values are provided in the Supporting Information (Table S14). As reported previously,\cite{Caricatotransitionlrccandeomcc} the dipole strength obtained from the EOM-CCSD method decreases almost linearly as the system size increases. If this trend is extrapolated, the dipole strength can even become negative, which is unphysical. This happens because the EOM-CC method is not strictly size-intensive for excited-state properties. This originates from the non-Hermitian nature of the similarity-transformed Hamiltonian, which leads to different left and right eigenvectors. Although both EOM-CC and LR-CC give the same eigenvalue, they differ in the formulation of the transition moment.   In LR-CC theory, transition moments retain their size-intensive nature, whereas size-intensivity is not preserved in EOM-CC.  For a single LiH molecule, the left and right transition moments obtained from both EOM-CC and LR-CC are nearly identical. However, when considering a system composed of multiple non-interacting LiH units, the two approaches differ. In EOM-CC, the right transition moment remains essentially unchanged with increasing system size, while the left transition moment decreases as the number of LiH molecules increases. Consequently, the overall transition moment in EOM-CC diminishes with system size, violating size intensivity. In contrast, LR-CC maintains consistent left and right transition moments regardless of the number of non-interacting subsystems, thereby preserving size-intensive behavior. The origin of the size dependence in EOM-CC can be traced to the disconnected nature of the ground-state left-vector amplitudes \textcolor{red}{.}\cite{PIECUCH1999295}. The qUCCSD method, being Hermitian, does not need any left vector and is free from this lack of size-intensivity, and the dipole strength remains almost unchanged with increasing system size.

\subsection{Oscillator Strengths}
\begin{table}[ht]
%\centering
\caption{Oscillator strength error statistics with respect to exFCI reference values using the aug-cc-pVTZ basis set. The sample size is $N=9$ dipole-allowed states.}
\label{tab:osc_errors_dz}
\begin{ruledtabular}
\begin{tabular}{lccccc}
Method  & ME & MAE & RMSE & STD & $\Delta_{\max}$ \\
\hline
ADC(3)     & 0.004 & 0.006 & 0.008 & 0.007 & 0.020 \\
CCSD       & 0.001 & 0.002 & 0.003 & 0.003 & 0.007 \\
%UCC3(2) & 9.000 & 0.002 & 0.003 & 0.004 & 0.004 & -0.005 & 0.011 \\
%qUCCSD(2/2) & 9.000 & -0.000 & 0.002 & 0.003 & 0.003 & -0.005 & 0.005 \\
qUCCSD(2)  & 0.000 & 0.002 & 0.003 & 0.003 & 0.005 \\
\end{tabular}
\end{ruledtabular}
\end{table}

Of the 15 excited states in Table~\ref{tab:all_dipole_tz}, nine are dipole-allowed from the ground state. Their length-gauge oscillator strengths are compared with the exFCI values in Table~\ref{tab:osc_errors_dz}. Both qUCCSD(2) and CCSD reproduce the exFCI oscillator strengths to within 0.007 for every transition, with an MAE of 0.002 and an RMSE of 0.003. The mean error of qUCCSD(2) is essentially zero, so the method shows no systematic tendency to over- or underestimate transition intensities. Its largest deviations occur for HCl ($-0.005$) and BF ($+0.005$), compared with a maximum deviation of 0.007 (BF) for CCSD. In relative terms, the BF error amounts to only about 1\% of the large oscillator strength of this state (0.468), whereas the HCl $^1\Pi$ intensity is underestimated by about 9\%, the largest relative deviation in the set. ADC(3) shows a positive ME (0.004) and an MAE three times larger than that of qUCCSD(2) (0.006 vs 0.002). Its largest errors occur for BF ($+0.020$) and BH ($+0.009$), two of the molecules for which its ground-state dipole moments also deviate most from exFCI.

Because qUCCSD(2) and ADC(3) evaluate the transition moments with the same second-order ISR expression [Eq.~\eqref{eq:spectral_amplitude}], their comparison isolates the effect of the quantities that enter this expression, namely the ground-state amplitudes and the excited-state eigenvectors $\mathbf{X}_n$. In ADC(3) both are obtained from perturbation theory, whereas in qUCCSD(2) they are taken from the iteratively converged qUCCSD amplitudes and the qUCCSD secular matrix. This replacement lowers the MAE from 0.006 to 0.002 and the maximum deviation from 0.020 to 0.005, in line with the improvement observed for the ground-state dipole moments. CCSD reaches a similar accuracy, but each transition moment requires both the left and right EOM eigenvectors together with the $\Lambda$ amplitudes. In qUCCSD(2), a single set of eigenvectors of a Hermitian secular matrix defines both sides of the transition moment, which is also the origin of the size-intensive behavior shown in Fig.~\ref{fig:size_extensivity}.

\subsection{Excited State Dipole Moments}
Table~\ref{tab:all_dipole_tz} also lists the dipole moments of all 15 excited states, classified as valence (V) or Rydberg (R).
The accuracy of qUCCSD(2) depends clearly on the character of the state. qUCCSD(2) gives more accurate dipole moments for valence states than for Rydberg states, with a mean absolute error of 0.052~D across the eight valence states and 0.091~D across the seven Rydberg states. Among the Rydberg states, the dipole moments of the $^1B_1$ and $^1A_2$ states of H$_2$O are overestimated by 0.120 and 0.144~D, respectively. The largest deviation occurs for the $^1A_2$ state of H$_2$CSi, for which qUCCSD(2) overestimates the dipole moment by 0.179~D. The diffuse electron densities of Rydberg states make their dipole moments particularly sensitive to the treatment of electron correlation and the flexibility of the basis set.
\begin{table}[ht]
\centering
\caption{Excited-state dipole moment error statistics (Debye) with respect to exFCI reference values using aug-cc-pVTZ basis set. The sample size (N) is 15.}
\label{tab:es_dip_errors}
\begin{ruledtabular}
\begin{tabular}{lccccc}
Method  & ME & MAE & RMSE & STD & $\Delta_{\max}$ \\
\hline
ADC(3)      & 0.034 & 0.090 & 0.118 & 0.113 & 0.297 \\
CCSD        & 0.013 & 0.048 & 0.060 & 0.059 & 0.119 \\
%qUCCSD(2)  & 0.026 & 0.071 & 0.085 & 0.081 & 0.181 \\
qUCCSD(2)   & 0.021 & 0.071 & 0.085 & 0.082 & 0.179 \\
\end{tabular}
\end{ruledtabular}
\end{table}

The error statistics in Table~\ref{tab:es_dip_errors} show small mean errors for all three methods (0.013--0.034~D), so none of them has a strong systematic bias; the methods differ mainly in the spread of the errors. CCSD is the most accurate (MAE 0.048~D, $\Delta_{\max}$ 0.119~D), followed by qUCCSD(2) (MAE 0.071~D, $\Delta_{\max}$ 0.179~D), while ADC(3) gives the largest errors (MAE 0.090~D, $\Delta_{\max}$ 0.297~D). Unlike the ground state, where qUCCSD(2) and CCSD perform equally well, a clear gap to CCSD remains for excited states. Nevertheless, qUCCSD(2) improves significantly on ADC(3).

\begin{table}[ht]
\centering
\caption{Combined ground- and excited-state dipole moment error statistics (Debye) 
with respect to exFCI reference values using the aug-cc-pVTZ basis set. The sample size (N) is 26.}
\label{tab:combined_dip_errors_tz}
\begin{ruledtabular}
\begin{tabular}{lccccc}
Method & ME & MAE & RMSE & STD  & $\Delta_{\max}$ \\
\hline
ADC(3)  & 0.060 & 0.099 & 0.134 & 0.120  & 0.297 \\
CCSD  & 0.018 & 0.045 & 0.056 & 0.053  & 0.119 \\
qUCCSD(2) & 0.019 & 0.056 & 0.071 & 0.069 & 0.179 \\
\end{tabular}
\end{ruledtabular}
\end{table}

The combined set of 11 ground-state and 15 excited-state dipole moments (Table~\ref{tab:combined_dip_errors_tz}) confirms these trends. The MAE of ADC(3) (0.099~D) is 2.2 times that of CCSD (0.045~D) and 1.8 times that of qUCCSD(2) (0.056~D), and its maximum deviation (0.297~D) is also the largest. For comparison, Dreuw and co-workers reported an MAE of 0.19~D for ADC(3/2) dipole moments.\cite{schneider2026mixed} qUCCSD(2) outperforms ADC(3) despite its ADC(2)-level density presumably for two reasons. The iteratively converged qUCCSD ground-state amplitudes replace the single perturbative step of ADC, and the qUCCSD secular matrix gives more accurate excited-state vectors. Compared with ADC(3), its ph--ph and pphh--ph/ph--pphh blocks contain additional higher-order contributions, while the pphh--pphh block is identical in both methods. Larger deviations from exFCI remain for Rydberg states and for systems with stronger multireference character, where higher excitations such as triples may be required.

\begin{figure}[t]
    \centering

    % -------- Second Row --------
    \begin{subfigure}[b]{0.4\textwidth}
        \centering
        \includegraphics[width=\linewidth]{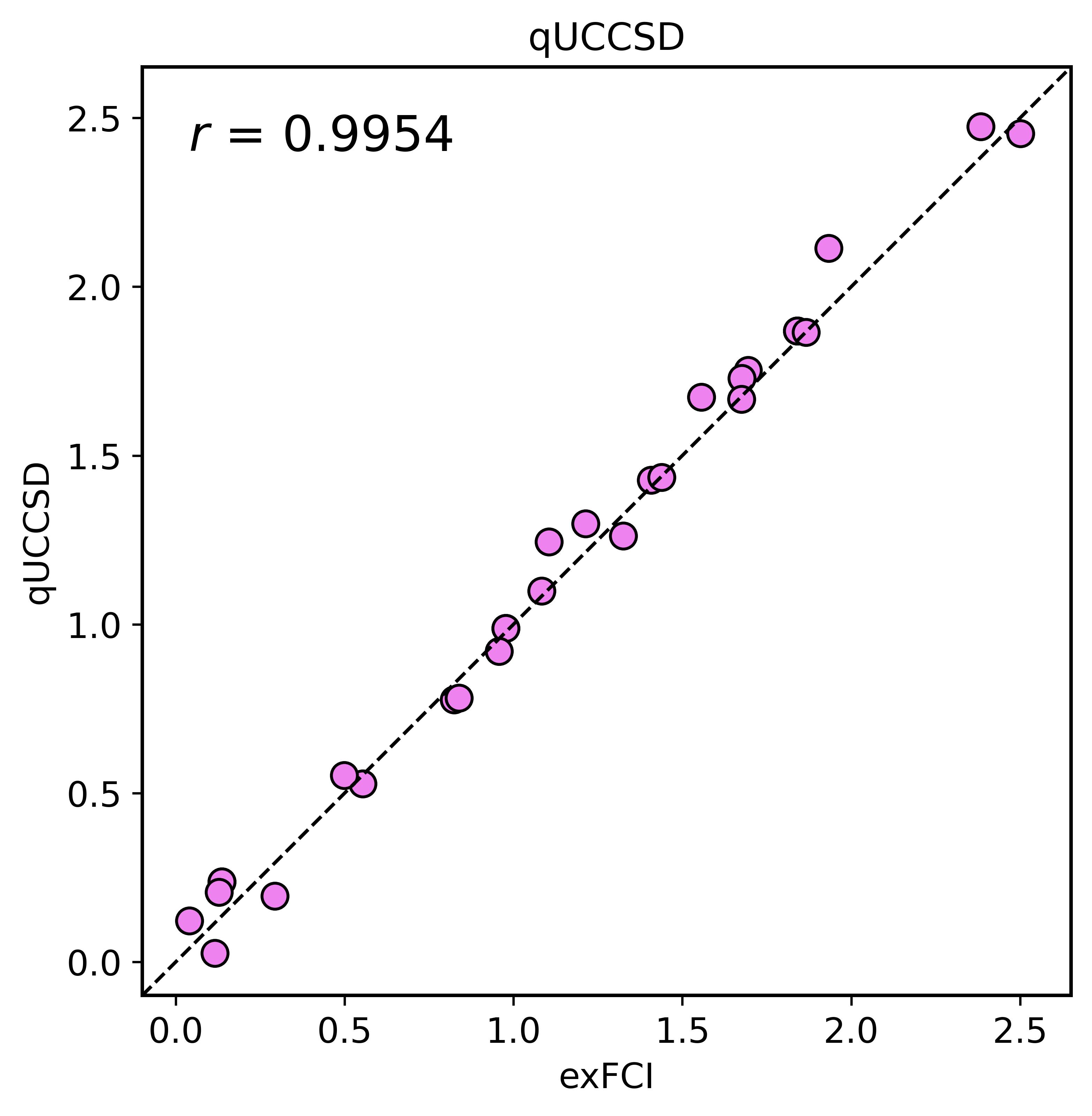}
        \caption{}
    \end{subfigure}
    %\hfill
    \begin{subfigure}[b]{0.4\textwidth}
        \centering
        \includegraphics[width=\linewidth]{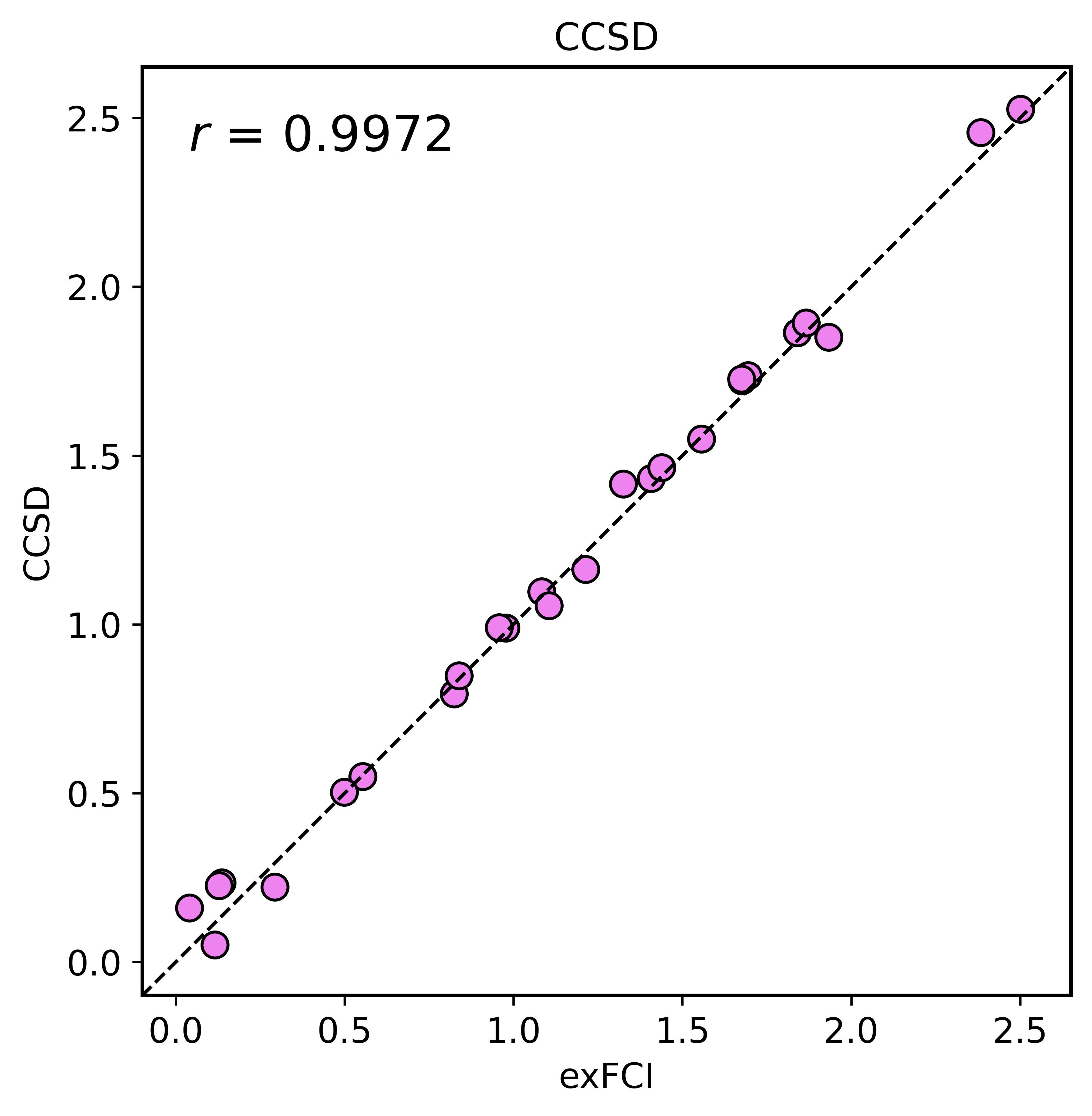}
        \caption{}
    \end{subfigure}

    \caption{
    Correlation between combined ground- and excited-state dipole moments
    and exFCI reference values.
    The dashed line represents the identity line.
    }
    \label{fig:gses_correlation}
\end{figure}

Figure~\ref{fig:gses_correlation} compares the combined qUCCSD(2) and CCSD dipole moments with exFCI. Both data sets lie close to the identity line over the full range of values, with correlation coefficients of $r=0.9954$ for qUCCSD(2) and $r=0.9972$ for CCSD, with no systematic deviation from the diagonal. qUCCSD(2) shows slightly more scatter than CCSD, but both are in good overall agreement with exFCI.

\section{Conclusions}
In this paper, we present an implementation of the second-order polarization propagator approximation to qUCCSD [qUCCSD(2)] for the calculation of excitation energies as well as ground- and excited-state properties. Both ground and excited-state properties, including oscillator strengths and dipole moments, are evaluated using a second-order ISR approximation. The implementation has been tested on a representative set of molecules exhibiting both valence and Rydberg-type excitations. The performance of qUCCSD is assessed through comparison with CCSD and ADC(3) results, using exFCI as the reference. Overall, qUCCSD demonstrates improved accuracy over ADC(3) for both ground- and excited-state properties and, in several cases, achieves results comparable to CCSD. The mean absolute errors of the qUCCSD(2) dipole moments are 0.035~D for the ground states and 0.071~D for the excited states, whereas CCSD yields slightly better agreement for excited states, particularly for those with pronounced Rydberg character. The ability of qUCCSD(2) to provide size-intensive transition moments without the need to compute left eigenvectors, combined with only a modest loss in accuracy, makes it an attractive alternative to EOM-CCSD for the simulation of UV-Vis spectra. Including higher excitations, such as triples, should improve the description of systems with significant multireference character or strong double-excitation contributions. Work in this direction is in progress.

\begin{acknowledgments}
The authors acknowledge support by the EU NextGenerationEU through the Recovery and Resilience Plan for Slovakia under project No. 09I03-03-V04-00117. This work was supported by the ANRF India (Project No. CRG/2023/002558).
\end{acknowledgments}
\section*{Supporting Information}
The Supporting Information contains the ground- and excited-state dipole moments along with their corresponding energy values, computed using different methods and basis sets. It also includes dipole-strength data for noninteracting LiH molecules.

\section*{References}
\bibliographystyle{aipnum4-1}
%\bibliography{references,paper,aipsamp}% Produces the bibliography via BibTeX.
\bibliography{reference1}

\end{document}